\documentclass {article}
  \usepackage{float}
  \usepackage{epsfig} 
  \begin{document}  
  \title {Non-Universality in Semi-Directed Barabasi-Albert Networks}
  \author {M.A. Sumour $^{1} $, \and M.A. Radwan$^{1} $} 
 \maketitle  

$^1$ Physics Department,
  Al-Aqsa University, P.O.4051, Gaza, Gaza Strip, Palestinian Authority, e-mail: msumoor@yahoo.com ,
   ma.radwan@yahoo.com .
 \section*{Abstract} 

In usual scale-free networks of Barab\'asi-Albert type, a newly added node selects randomly $m$ 
neighbors from the already  existing network nodes, proportionally to the number of links these had before.
 Then the number $N(k)$ of nodes with $k$ links  each decays as $1/k^{\gamma}$ where $\gamma=3$ is universal,
 i.e. independent of $m$. 
 Now we use a limited directedness in the construction of the network, as a result of which the exponent $\gamma$
  decreases from 3 to 2 for increasing $m$.

  \section*{Keywords}

 General BA, directed BA, undirected BA, Neighbors, and Fortran programs 

 \section*{Introduction:} 

 The Barab\'asi-Albert network is growing such that the probability of a new site to be connected
 to one of the already existing sites,   is proportional to the number of previous connections to this 
already existing site: The  rich gets richer, in this way, each new site selects  
exactly $m$ old sites as neighbors.

 In directed Barab\'asi-Albert networks[1], the network itself was produced in the standard way, but 
then the neighbor relations were  such that if A has B  as a neighbor, B in general does not
 have A as a neighbor[2]. 

The undirected Barab\'asi-Albert network [3], usually is grown in the same way, but then the 
neighbor relations were such that if A has B as  a neighbor, B has A as a neighbor[2].
 
  Now we  try to introduce a directedness already in the construction of the network[3]. 
 In the undirected Barab\'asi-Albert network [4-5], if a new node selects $m$ old nodes as neighbor,
 then the $m$ old nodes are added to the  Kert\'esz list, and the new node is also added $m$ times to that list.
 
 Connections are made with $m$ randomly selected elements of that list. 
 If one would only add the old nodes to the list, then only the  initial core can be selected as neighbors, 
which is not interesting.
  But if one adds the $m$ old nodes plus only once (and not $m$ times) the new node, then one has 
a semi-directed network.  In this modified Barab\'asi-Albert model one can (but we don't do that here) 
put  in neighbor relations which are directed or which are undirected.

  \section*{Data and Simulation:} 

 We use the FORTRAN program[6], as in appendix, with different $ m =1,2,3,\dots,10$, and the maxtime = 41,000,000 nodes.


     \begin{figure}[H]
       \centerline { 
        \epsfig{figure=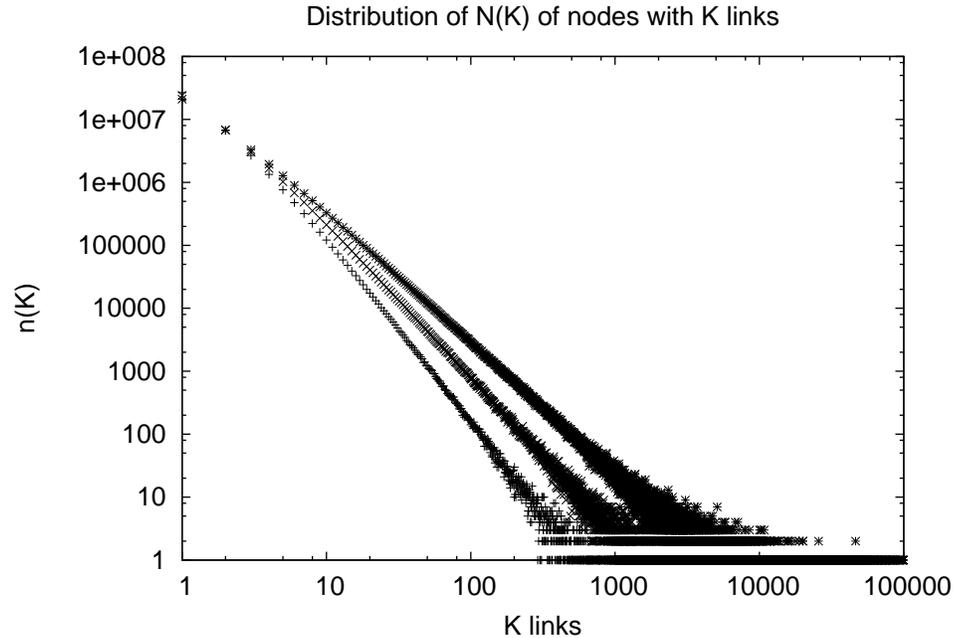, height=5.0in , angle=270}
         }
         \caption{ $n(k)$ versus $k$ links at number of neighbors $m= 1, 2,$ and 10, from left to right,   
         maxtime=40000000  in log-log plot.}
        \label{Fig1} 
  \end{figure}

   Figure(1) shows the number $n(k)$ of nodes with $k$ links each.
 Changes are seen with increasing number of neighbors $m$,  and the curvature shows that the exponent
 should be determined from the region on large $k$ only. 

We are going to plot the slopes versus $1/m$ to get the  second  plot , which makes clearer the possible 
extrapolation towards  infinite $m (m=\infty, 1/m=0)$ as shown in figure 2.

 \begin{figure}[H] 
   \centerline {
     \epsfig{figure=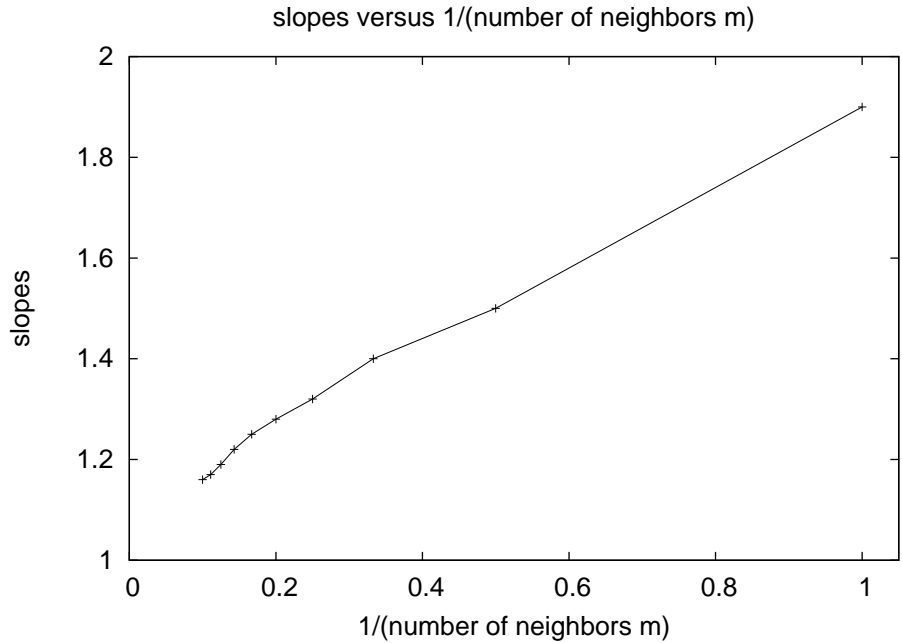, height=5.0in , angle=270}
      }  
     \caption{Slopes versus $1/m$ for number of neighbors $m= 1,2,3,\dots,10$, maxtime=41000000.}
      \label{Fig2}
  \end{figure}

   \section*{Conclusions:}

 The exponent gamma for the decay of the number $ n(k)$ of nodes with $k$ links each, changes with increasing 
number of neighbors $m$,  the exponent should be determined from the region on large $k$ only.
 
 Fig.2 shows $\gamma-1$ since we summed all $k$ into bins limited by powers of two, e.g, $4,5,6,7$ is one of 
the many points onto which we  fitted the exponent $\gamma-1$.

   \section*{Acknowledgements:}   

The authors would like to thank Prof. Stauffer  for many valuable suggestions, fruitful discussions and 
constructive advice  during the development of this work. 

 \newpage
 \section*{Appendix:} 

  {\small \begin{verbatim}  
      parameter(nrun=1 ,maxtime=40000000 ,m=2,iseed=1,max=maxtime+m,
     1      length=1+(1+m)*maxtime+m*(m-1))
       integer*8 ibm 
       real*8 factor
       dimension k(max), nk(131070), list(length), nklog(30)
       data nk/131070*0/,nklog/30*0/
       OPEN (UNIT=60,FILE='N(K)M2.DAT')
       WRITE (60,*)'#(nrun, maxtime, m, iseed)',nrun, maxtime, m,iseed
       ibm=2*iseed-1   
       factor=(0.25/2147483648.0d0)/2147483648.0d0 
       fac=1.0/0.69314 
c      factor=0.5/2147483648.0d0
       do 5 irun=1,nrun 
        do 3 i=1,m  
         do 7 j=(i-1)*(m-1)+1,(i-1)*(m-1)+m-1
 7           list(j)=i 
 3       k(i)=m-1  
         L=m*(m-1) 
         if(m.eq.1) then 
          L=1     
          list(1)=1
         endif 
c       All m initial sites are connected with each other
        do 1 n=m+1,max 
         do 2 new=1,m 
 4        ibm=ibm*16807 
          j=1+(ibm*factor+0.5)*L  
          if(j.le.0.or.j.gt.L) goto 4 
          j=list(j)  
          list(L+new)=j  
 2        k(j)=k(j)+1 
         list(L+m+1)=n 
         L=L+m+1 
 1       k(n)=1  
         WRITE (60,*), '# (irun )', irun 
C       print *,'# (irun )', irun
        do 5 i=1,max
         k(i)=min0(k(i),131070)
 5       nk(k(i))=nk(k(i))+1    
       do 6 i=1,131070 
 6      if(nk(i).gt.0) write (60,*), i,nk(i) 
       do 9 i=1,131070      
        j=alog(float(i))*fac  
 9      nklog(j)=nklog(j)+nk(i) 
       do 10 j=0,18
 10     WRITE (60,*), sqrt(2.0)*2**j,nklog(j),j 
       stop      
       end
 \end{verbatim} }
 
  \section*{References:} 
\parindent 0pt

 [1] R. Albert and A. L. Barab\'asi, ,,Statistical Mechanics of Complex Networks.
     '' Reviews of Modern  Physics, Physics 74, 47 (2002).
  
\medskip 
 [2] David P. Landau, Kurt Binder, A Guide to Monte Carlo, Simulation in Statistical Physics.
  Cambridge University Press; (2002).

\medskip 
 [3]A.L.Barab\'asi, Linked, Perseus, Cambridge (2002) and Science 325, 412 (2009). 
There are other network  articles in that part of the Science issue (24 July 2009).

\medskip 
 [4]M. A. Sumour, M. M. Shabat , Int. M. Phys. C, 16, 585(2005).
 
\medskip 
 [5] M. A. Sumour, M. M. Shabat  and D. Stauffer, Islamic University Journal  
(Series of Natural Studies and Engineering) 14, 209 (2006), arXiv:cond-mat/0504460.

\medskip 
 [6] D. Stauffer, Random networks and small worlds, preprint for: Handbook Modelling and Simulation in the
 Social Sciences, to be edited by N. Braun and N.J.  Saam, VS-Verlag, Wiesbaden (in German language).

    \end{document}